\documentclass{article}
\usepackage{spconf,amsmath,graphicx,hyperref}
\usepackage{booktabs}
\usepackage{multirow}
\usepackage{adjustbox}
\usepackage{xcolor}
\usepackage{colortbl}
\usepackage{array}
\usepackage{makecell}
\usepackage{graphicx}
\usepackage{tabularx}
\usepackage{multirow}
\usepackage{float}
\usepackage{booktabs}
\usepackage{amsfonts} 
\usepackage[table]{xcolor}
\usepackage{threeparttable}
\usepackage{pifont}
\usepackage{url}
\definecolor{bestcolor}{HTML}{E8F5E9}
\definecolor{headercolor}{HTML}{F5F5F5}
\definecolor{sectioncolor}{HTML}{FAFAFA}
\definecolor{groupA}{RGB}{235, 243, 250}  
\definecolor{groupB}{RGB}{250, 245, 235}  

\newcolumntype{C}[1]{>{\centering\arraybackslash}p{#1}}
\newcolumntype{R}[1]{>{\raggedleft\arraybackslash}p{#1}}

\definecolor{flomssBlock}{HTML}{EAF3FF}
\definecolor{flomssBlockDark}{HTML}{D8E9FF}
\definecolor{gainGreen}{HTML}{008A2E}
\definecolor{lossOrange}{HTML}{B35C00}

\ninept

\usepackage{xcolor}

\title{FLOMSS: GENERATIVE MUSIC SOURCE SEPARATION BY FLOW MATCHING}
 \name{
 \begin{tabular}{@{}c@{}}
 Shuyang Cui$^{1}$,
 \quad Zhi Zhong$^{1}$,
  \quad Keisuke Toyama$^{1}$ \\
 \quad Koichi Saito$^{1}$,
 \quad Qiyu Wu$^{1}$,
 \quad Yukara Ikemiya$^{1}$,
 \quad Junghyun Koo$^{1}$,
 \quad Kin Wai Cheuk$^{1}$,\\
 Woosung Choi$^{1}$,
 \; Giorgio Fabbro$^{1}$,
 \; Wei-Hsiang Liao$^{1}$,
 \; Shusuke Takahashi$^{1}$,
 \; Yuki Mitsufuji$^{1}$
 \end{tabular}}
 \address{$^{1}$Sony Group Corporation \;
 }

\begin{document}

\maketitle

\begin{abstract}
Music source separation (MSS) has been conventionally addressed by discriminative methods. 
Recently, flow matching has emerged as a strong generative paradigm for modeling complex data distributions,
with successful applications across diverse domains. 
These advances motivate us to systematically investigate flow matching for MSS
in a four-source (Vocals, Bass, Drums, Other) setting in full-band stereo. 
Specifically, we investigate prediction targets, flow paths, single-source versus multi-source modeling, and mixture consistency constraints. 
Beyond the signal-level metrics commonly used for MSS, such as signal-to-distortion ratio, 
we further evaluate MSS methods using embedding-based metrics, 
as well as a listening test, providing a comprehensive benchmark.
On MUSDB18-HQ, our proposed model exhibits competitive performance in both objective and subjective tests, highlighting the potential of flow matching for generative MSS.
Demo: https://ik4sumii.github.io/FLOMSS/


\end{abstract}

\vspace{-2mm}
\section{Introduction}\label{sec:introduction}
\vspace{-2mm}
Music source separation (MSS) ~\cite{araki202530yearssourceseparation} involves reconstructing individual source signals from a music mixture. 
Deep learning research in MSS has been dominated by models trained with discriminative objectives, 
achieving strong performance~\cite{lu2023musicsourceseparationbandsplit, rouard2022hybridtransformersmusicsource,Tong2024SCNetSC}. 
Meanwhile, generative modeling has received increasing attention in related fields, such as in speech enhancement~\cite{Richter_2023}, speech separation~\cite{scheibler2025sourceseparationflowmatching}, and music bandwidth extension~\cite{moliner2022behmgan}. 
In MSS, recent generative methods primarily use learned source distributions for prior-based inference or conditional inpainting but generally do not explicitly learn mixture-to-source generative mapping. 
For example, InstGlow~\cite{Zhu_2022} optimizes latent variables under independent source priors and a mixture likelihood;
Multi-source Diffusion (MSDM)~\cite{mariani2024multisourcediffusionmodelssimultaneous} learns the joint source prior and performs separation via posterior sampling under the mixture constraint;
MGE-LDM~\cite{chae2025mgeldmjointlatentdiffusion} extends the multi-source setting, formulating source separation as a conditional inpainting in the latent space.
Although these works establish foundations for generative MSS, they do not produce full-band stereo audio, which is essential to high-fidelity music production.

More recently, flow matching (FM)~\cite{lipman2023flowmatchinggenerativemodeling} has emerged as a compelling alternative to diffusion for generative modeling. Unlike conventional diffusion-based training, flow matching learns continuous transports by regressing velocity fields along predefined probability paths. Optimal-transport or rectified paths can offer straighter trajectories, faster sampling, and strong generalization~\cite{tong2024improvinggeneralizingflowbasedgenerative}. 
These advantages have begun to transfer to audio separation: flow-matching models have been introduced for language-queried target sound separation~\cite{yuan2025flowseplanguagequeriedsoundseparation}, speech enhancement~\cite{flowse, wang2026rethinkingflowdiffusionbridge}, and speech source separation~\cite{scheibler2025sourceseparationflowmatching}. More recently, SAM Audio~\cite{shi2025samaudiosegmentaudio} conditions a flow-matching model on text, visual, or temporal prompts for open-domain audio separation, including musical instrument extraction.
However, flow matching has not been systematically studied and evaluated for MSS, which targets predefined musical stems in full-band stereo mixtures~\cite{araki202530yearssourceseparation,lu2023musicsourceseparationbandsplit,rouard2022hybridtransformersmusicsource}, unlike common speech-separation settings that separate multiple speakers from monaural mixtures~\cite{scheibler2025sourceseparationflowmatching}.

In this paper, we study supervised conditional flow matching (CFM) models for full-band stereo MSS, which separate a song into four sources: \textit{Vocals}, \textit{Bass}, \textit{Drums}, and \textit{Other}, where the \textit{Other} source contains all remaining sources. We train and evaluate our models on MUSDB18-HQ~\cite{Rafii2019MUSDB18HQA}. We further extend the evaluation beyond signal-to-distortion ratio (SDR) by introducing spectral and embedding-based metrics based on pretrained semantic audio encoders and benchmark our models against prior discriminative and generative methods under these metrics.




Our main contributions are as follows: 
(1) We introduce \textit{FLOMSS}, an open-source\footnote{Will be open upon the acceptance} generative framework for full-band stereo MSS; 
(2) We present a systematic study of flow matching for MSS, covering prediction targets, flow paths, single- and multi-source modeling, and mixture consistency strategies; 
(3) We introduce a comprehensive evaluation for generative MSS covering signal-based and embedding-based metrics; 
(4) We show that FLOMSS surpasses prior generative baselines and improves spectral and embedding-based metrics over a same-backbone discriminative baseline while achieving competitive performance compared with state-of-the-art discriminative methods.

\vspace{-2mm}
\section{Background}\label{sec:background}
\vspace{-2mm}

\subsection{Music Source Separation}\label{sec:music_source_separation}
\label{ssec:back_mss}
\vspace{-1mm}
Let $x_k\in\mathbb{R}^{\mathrm{ch}\times N}$ denote the waveform of source $k$ with $\mathrm{ch}=2$ for stereo audio, and let
$
    m=\sum_{k=1}^{K}x_k
$
be the observed mixture, where $\mathrm{ch}$ and $N$ denote the numbers of
channels and samples, respectively.
A typical supervised formulation of MSS learns either $K$ source-specific predictors $\mathrm{\mathbf{DNN}}_{\theta,k}(m)$ or a joint predictor
$\mathrm{DNN}_\theta(m) = (\hat{x}_1,\dots,\hat{x}_K)$,
using reconstruction objectives, \emph{e.g.}, waveform-domain $\ell_1$ loss, a magnitude- or complex-valued Short-time Fourier Transform (STFT)
reconstruction loss.
This formulation covers both time--frequency masking separators~\cite{stoter19} and waveform-to-waveform separators~\cite{defossez2021musicsourceseparationwaveform}.

\vspace{-2mm}
\subsection{Flow Matching}\label{sec:flow_matching}
\vspace{-1mm}
FM trains a 
continuous normalizing flow~\cite{chen2019neuralordinarydifferentialequations} by regressing the velocity field of a prescribed probability path~\cite{lipman2023flowmatchinggenerativemodeling}.
Since the marginal target velocity is generally intractable, conditional flow matching (CFM) instead regresses tractable sample-conditioned velocities with the same gradient as the marginal FM objective~\cite{lipman2023flowmatchinggenerativemodeling}. Let $\mathbf{x}_t$ denote the state at time $t$,
$\mathbf{x}_1$ a target sample, and $\mathbf{c}$ an observed
condition. 
For $t\in(0,1)$ with $\sigma_t>0$, a Gaussian ($\mathcal{N}$) conditional probability path ($p$) and its corresponding velocity field are
\begin{equation}
\label{eq:flow-matching}
\vspace{-1mm}
\begin{aligned}
p_t(\mathbf{x}_t\mid\mathbf{x}_1,\mathbf{c})
 &= \mathcal{N}(\mathbf{x}_t;
    \boldsymbol{\mu}_t,\sigma_t^2\mathbf{I}),\\
\vspace{-1mm}
\mathbf{u}_t(\mathbf{x}_t\mid\mathbf{x}_1,\mathbf{c})
 &= \dot{\boldsymbol{\mu}}_t
  + \frac{\dot{\sigma}_t}{\sigma_t}
    (\mathbf{x}_t-\boldsymbol{\mu}_t),
\end{aligned}
\vspace{-1mm}
\end{equation}
where $\boldsymbol{\mu}_t$ is the prescribed mean trajectory, $\sigma_t^2$ is the variance schedule, $\mathbf{I}$ is the identity matrix, and $\dot{(\cdot)}$ denotes differentiation with respect to $t$.

The training of a CFM can be formulated with either a \textit{velocity-prediction} ($\mathbf{v}$-prediction), which estimates the target velocity, or a \textit{data-prediction} ($\mathbf{x}$-prediction), which estimates $\mathbf{x}_1$. These objectives provide alternative parameterizations of the same prescribed probability path but can induce different optimization behavior under finite model capacity and finite data~\cite{karras2022elucidatingdesignspacediffusionbased,yang2025investigatingtrainingobjectiveflow}. 
We consider two formulations: optimal-transport-inspired CFM (OT-CFM)
~\cite{tong2024improvinggeneralizingflowbasedgenerative, flowse, welker2025flowdecflowbasedfullbandgeneral}, 
and Schr\"odinger-bridge-inspired CFM (SB-CFM)~\cite{tong2024simulationfreeschrodingerbridgesscore}. The latter uses Gaussian conditional paths with fixed endpoints and nonzero intermediate variance.
Implementation details are available in our open-source repository.

\vspace{-2mm}
\section{Music Source Separation by Conditional Flow Matching}\label{sec:proposed_method}
\vspace{-2mm}
\subsection{Problem Definition}\label{sec:problem_definition}
\vspace{-1mm}
Following~\cite{Richter_2023}, we define CFM in the compressed complex-STFT domain. The magnitude dynamic range is compressed element-wise by a power law
\begin{equation}
\label{eq:power-law}
\widetilde X = \beta |X|^\alpha e^{j\angle X},
\end{equation}
where $X$ is the complex spectrogram, 
while $\alpha$ and $\beta$ denote the compression factor and global scaling factor, respectively.
Let $M$ and $X_k$ denote the mixture and source-$k$ spectrogram, respectively, and let $\widetilde M$ and $\widetilde X_k$ denote their compressed representations obtained using Eq.~\ref{eq:power-law}. 
The compression produces more consistently scaled inputs relative to the Gaussian perturbations used in CFM~\cite{Richter_2023}. At inference, we apply inverse compression and inverse STFT to the estimated source spectrogram to reconstruct the waveform.

We then formulate MSS as conditional generation: given an observed mixture $\widetilde M$ and a target source \(k\), the model estimates the conditional distribution \(p(\widetilde X_k\mid \widetilde M)\). 

Building on CFM introduced in Sec.~\ref{sec:flow_matching}, we define the Gaussian conditional probability path in Eq.~\ref{eq:flow-matching} as
\begin{equation}
p_t(\mathbf{x} \mid \mathbf{x}_1, \widetilde M)
=
\mathcal{N}\!\bigl(\mathbf{x}; \boldsymbol{\mu}_t(\mathbf{x}_1,\widetilde M), \sigma_t^2 \mathbf{I} \bigr),
\end{equation}
where
$\mathbf{x}_1=\widetilde X_k$ and 
$\boldsymbol{\mu}_t = (1-t)\widetilde M + t\mathbf{x}_1$.
For each mixture--source pair, the conditional probability path transports samples from $\mathcal{N}(\widetilde M,\sigma_0^2\mathbf I)$ to $\mathcal{N}(\widetilde X_k,\sigma_1^2\mathbf I)$.
We train a mixture-conditioned model to estimate the corresponding time-dependent velocity field via CFM.

\vspace{-1mm}
\subsection{Single-source Modeling}
\label{ssec:single-source modeling}
\vspace{-1mm}
We first consider a \emph{single-source} setting, in which a model is solely trained for one target source $k$ to estimate $\widetilde X_k$ from $\widetilde M$.

\noindent\textbf{Probability flow path.} We investigate both the OT-CFM path and the SB-CFM path introduced in Sec.~\ref{sec:flow_matching}. Both paths share the same linear mean interpolation but differ in their variance schedule $\sigma_t^2$. 

\noindent\textbf{Prediction parameterization.}
For $\mathbf{v}$-prediction, the network estimates the velocity field and minimizes 
$\mathcal{L}_{\mathrm{\mathbf{v}}}=\mathbb E[\|\mathrm{\mathbf{DNN}}_\theta(\mathbf{x}_t,t,\widetilde M)-\mathbf u_t(\mathbf{x}_t\mid\mathbf{x}_1,\widetilde M)\|_2^2]$, where the expectation is over training pairs, time $t$, and path samples $\mathbf{x}_t$. 
For $\mathbf{x}$-prediction, the network estimates the target sample and minimizes 
$\mathcal{L}_{\mathrm{\mathbf{x}}}=\mathbb E[\|\mathrm{\mathbf{DNN}}_\theta(\mathbf{x}_t,t,\widetilde M)-\mathbf{x}_1\|_2^2]$. 
\noindent\textbf{Discriminative baseline.}
We include a discriminative baseline using the same backbone to investigate the differences brought by generative modeling. 
It directly predicts the target sample from the mixture and minimizes
$\mathcal L_{\mathrm{Disc.}}
=\mathbb E[\|\mathrm{\mathbf{DNN}}_\theta(\widetilde M)-\widetilde X_k\|_2^2]$.
\vspace{-1mm}
\subsection{Multi-source Modeling}
\label{ssec:multi-source modeling}
\vspace{-1mm}

We extend single-source CFM to jointly predict all $K$ sources with only one model. 
To investigate the impact of this change, we keep the same backbone as the single-source models, 
with only the input and output channels expanded to process the concatenated four source states. 
The output channels follow a fixed source order, and the multi-source variants differ only in the flow path domain and mixture-consistency mechanism. 


We use OT-CFM with $\mathbf{x}$-prediction in this joint source space, as this setting shows a balance across various metrics, which is shown in Table~\ref{tab_ablation_all_stems} and will be discussed in Sec.~\ref{5.1}.


It is intuitive that the summation of sources should be identical to the input mixture, which is called \textit{mixture consistency} (MC). 
Introducing constraints based on MC when training source separation models has shown promise~\cite{scheibler2025sourceseparationflowmatching, sawata2021all}. 
To this end, we explore two ways of injecting MC-based constraints into FLOMSS.

\noindent\textbf{Soft mixture consistency.}\label{method:softmc}
On top of the unconstrained multi-source FLOMSS setting, we introduce an MSE loss between the summation of estimated sources and the input, 
making the total loss
\begin{equation}
\label{eq:softmc}
\vspace{-1.5mm}
    \mathcal{L}_\mathrm{SoftMC}=\mathcal{L}_{\mathrm{\mathbf{x}}}+0.1\mathcal{L}_\mathrm{MSE},
\vspace{-0.5mm}
\end{equation}
which encourages \textit{soft} mixture consistency without constraining the flow path.

\noindent\textbf{Hard mixture consistency.}
Unlike Soft MC, hard MC enforces exact mixture consistency throughout the probability path~\cite{scheibler2025sourceseparationflowmatching}.  
We apply this constraint to full-band stereo MSS. Since compression in Eq.~\ref{eq:power-law} is nonlinear and does not preserve source additivity, we define the constrained paths in the waveform or complex-STFT domain.
Let $\mathbf{x}_{t,\mathrm{w}}^{(k)}$ and $\mathbf{x}_{t,\mathrm{s}}^{(k)}$ denote the corresponding states of source $k$ at time $t$ . Following~\cite{scheibler2025sourceseparationflowmatching}, we require $\sum_{k=1}^{K}\mathbf{x}_{t,\mathrm{w}}^{(k)}=m$ or $\sum_{k=1}^{K}\mathbf{x}_{t,\mathrm{s}}^{(k)}=M$, respectively, at all times $t$, where $m$ is the mixture waveform and $M$ its complex spectrogram. We evaluate both path domains while retaining the compressed complex-STFT representation for the backbone.

\begin{table}[t]
\centering
\vspace{-4mm}
\vspace{4pt}

\caption{Ablation study on flow paths and training targets (average across all stems). 
The predictive baseline uses the same model architecture as described in Sec.\ref{ssec:single-source modeling}. 
\textbf{Bold} indicates the best result among ablations, \underline{underline} indicates the second best. $\downarrow$ indicates lower is better, $\uparrow$ indicates higher is better.}
\label{tab_ablation_all_stems}

\begin{adjustbox}{max width=\columnwidth}
\renewcommand{\arraystretch}{1.15}
\setlength{\tabcolsep}{4pt}
\begin{tabular}{
    l l
    !{\color{gray!40}\vrule}
    c c c
    !{\color{gray!40}\vrule}
    c
    !{\color{gray!40}\vrule}
    c
}
\toprule
& &
\multicolumn{3}{c!{\color{gray!40}\vrule}}{\cellcolor{groupA}\textit{Ablations}} &
\cellcolor{groupB}\textit{Baseline} &
\cellcolor{groupA}\textit{Ref.} \\
\cmidrule(lr){3-5} \cmidrule(lr){6-6} \cmidrule(lr){7-7}
\textbf{Group} & \textbf{Metric} &
\textbf{OT-CFM-}$\mathbf{v}$ & \textbf{OT-CFM-}$\mathbf{x}$ & \textbf{SB-CFM-}$\mathbf{x}$ &
\textbf{Discriminative} & \textbf{Mixture} \\
\midrule

\multirow{4}{*}{\textbf{Signal}}
& Mel $\downarrow$            & \underline{1.71} & \textbf{1.63} & 2.99            & 2.30  & 4.71 \\
& STFT $\downarrow$           & \underline{2.81} & \textbf{2.69} & 3.72            & 3.16  & 6.13 \\
& Wav. $\downarrow$           & 0.0241           & 0.0226     & \underline{0.0203} & \textbf{0.0197} & 0.0723 \\
& SDR \textbf{$\uparrow$}             & 3.23             & 3.77       & \underline{4.63}   & \textbf{4.95}  & -5.56 \\
\midrule

\multirow{5}{*}{\textbf{FD} $\downarrow$}
& VGG     & \textbf{0.77}    & \underline{0.87}     & 2.88   & 2.55   & 6.46   \\
& PaSST   & \textbf{225}  & \underline{261}   & 306 & 308    & 802    \\
& OpenL3  & \textbf{108} & \underline{116}        & 144   & 144   & 166  \\
& CLAP    & \textbf{0.23}    & \underline{0.31}     & 0.33   & 0.33   & 0.68   \\
& MERT    & \textbf{5.35}    & \underline{5.43}     & 6.87   & 6.23   & 13.8  \\
\midrule

\multirow{5}{*}{\textbf{MMD} $\downarrow$}
& VGG     & \textbf{2.66}    & \underline{3.26}     & 9.68   & 9.14   & 28.2  \\
& PaSST   & \textbf{36.4}    & \underline{44.5}     & 72.3   & 51.7   & 49.2   \\
& OpenL3  & \textbf{51.1} & \underline{59.4}        & 112   & 109   & 169   \\
& CLAP    & \textbf{0.98}    & \underline{1.58}     & 1.73   & 1.76   & 3.55   \\
& MERT    & \textbf{11.1} & \textbf{11.1}           & 21.7  & \underline{15.4}   & 51.5   \\
\midrule

\textbf{LPAPS} $\downarrow$ & CLAP
& \textbf{2.43}    & \underline{2.45}     & 2.72   & 2.58   & 4.13 \\

\bottomrule
\end{tabular}
\end{adjustbox}
\vspace{-3mm}

\vspace{-1mm}
\end{table}

\begin{figure}[tb]
    \centering
    \includegraphics[width=\columnwidth]{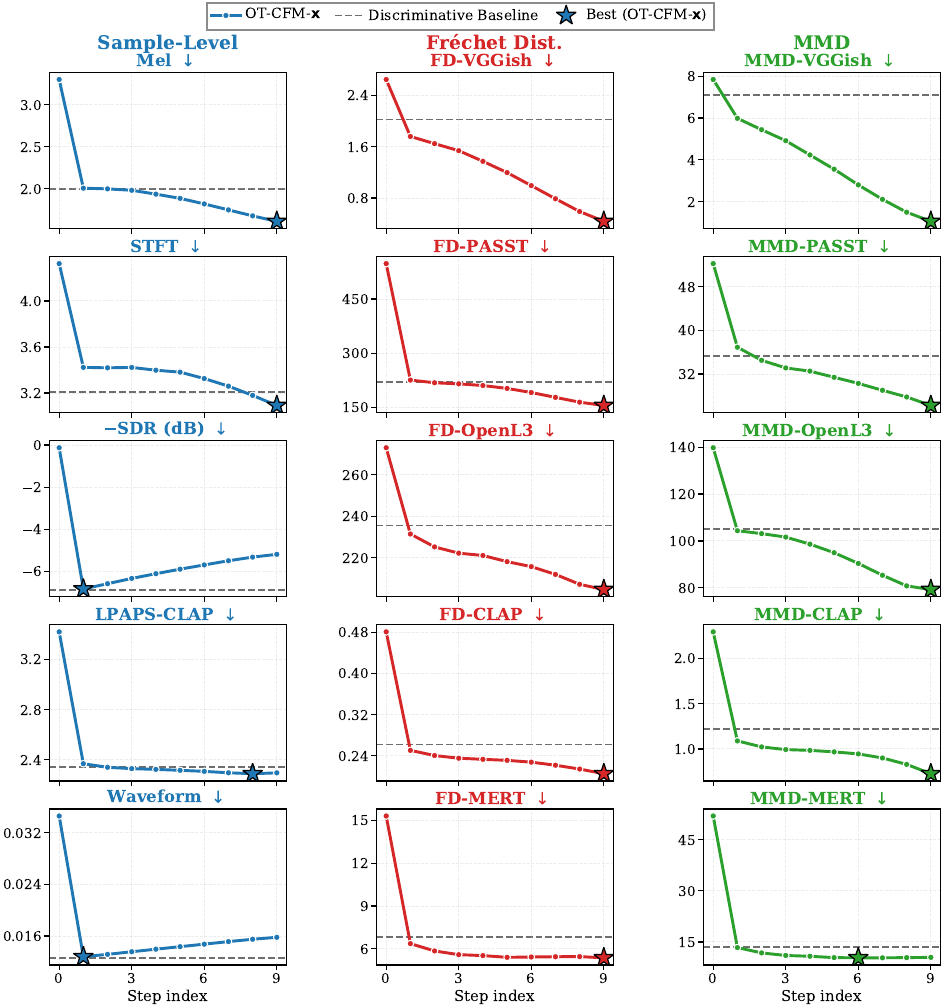}
    \vspace{-5mm}
    \caption{
        Stepwise behavior of OT-CFM-$\mathbf{x}$ for Vocals. 
Each subplot reports a value computed from the target prediction at the Midpoint ODE solver step of $i=0,\ldots,9$.
        Dashed gray lines indicate the discriminative baseline. 
        \ding{72} marks the best-performing position for each metric. 
        As the negative SDR is used, all metrics are reported as lower-is-better ($\downarrow$).
    }
    \label{fig:midpoint_sampler}
    \vspace{-5mm}
\end{figure}

\begin{table}[t]
\centering
\footnotesize

\caption{Comparison of MSS modeling approaches using average results across four stems.}
\label{tab_modeling}

\begin{adjustbox}{max width=\columnwidth}
\renewcommand{\arraystretch}{0.95}
\setlength{\tabcolsep}{2.5pt}
\begin{tabular}{
    l
    l
    !{\color{gray!40}\vrule}
    c c c c c c c
}
\toprule
\textbf{Method} &
\makecell[l]{\textbf{Path}\\\textbf{Domain}} &
\rotatebox{55}{\textbf{Mel}$\downarrow$} &
\rotatebox{55}{\textbf{SDR}$\uparrow$} &
\rotatebox{55}{\textbf{FD\textsubscript{VGG}}$\downarrow$} &
\rotatebox{55}{\textbf{FD\textsubscript{PaSST}}$\downarrow$} &
\rotatebox{55}{\textbf{FD\textsubscript{CLAP}}$\downarrow$} &
\rotatebox{55}{\textbf{FD\textsubscript{MERT}}$\downarrow$} &
\rotatebox{55}{\textbf{LPAPS\textsubscript{CLAP}}$\downarrow$} \\
\midrule

\rowcolor{headercolor}
\multicolumn{9}{l}{\textit{Single Source}} \\

\makecell[l]{OT-CFM-$\mathbf{x}$} &
\makecell[l]{Compressed STFT} &
\textbf{1.63} &
\textbf{3.77} &
\textbf{0.87} &
\textbf{261} &
\textbf{0.31} &
\textbf{5.43} &
\textbf{2.45} \\

\midrule
\rowcolor{headercolor}
\multicolumn{9}{l}{\textit{Multi Source}} \\

\makecell[l]{w/o MC} &
\multirow{2}{*}{\makecell[l]{Compressed STFT}} &
\underline{1.71} &
3.07 &
\underline{2.30} &
\underline{416} &
\underline{0.45} &
10.6 &
\underline{2.71} \\

\makecell[l]{Soft MC} &
&
1.73 &
\underline{3.69} &
2.87 &
446 &
0.47 &
13.7 &
\underline{2.71} \\

\cmidrule(lr){1-9}

\multirow{2}{*}{\makecell[l]{Hard MC}} &
\makecell[l]{Waveform} &
2.74 &
2.11 &
5.17 &
497 &
\underline{0.45} &
9.61 &
3.23 \\

&
\makecell[l]{STFT} &
2.71 &
2.34 &
6.30 &
597 &
0.55 &
\underline{8.83}     &
3.40 \\

\bottomrule
\end{tabular}
\end{adjustbox}
\vspace{-3mm}

\vspace{-5mm}
\end{table}

\begin{table*}[t]
\centering
\setlength{\abovecaptionskip}{0pt}
\setlength{\belowcaptionskip}{2pt}
\caption{Benchmark of music source separation (MSS) methods on the MUSDB18-HQ test set. 
SAM-Audio and MGE-LDM are text-conditioned models, which we found difficult to properly separate the ``other'' stem.
All other methods are evaluated across four stems and their average. 
\textit{Italic} names denote models trained solely on the MUSDB18-HQ training set. For single-channel models, we process the two mixture channels independently and stack the resulting estimates to obtain stereo outputs.}
\label{tab_benchmark}

\begin{adjustbox}{max width=\textwidth}
\renewcommand{\arraystretch}{1.2}
\setlength{\tabcolsep}{3pt}
\begin{tabular}{
    l
    !{\color{gray!40}\vrule}
    c c c c c c c
    !{\color{gray!40}\vrule}
    c c c c c c c
    !{\color{gray!40}\vrule}
    c c c c c c c
    !{\color{gray!40}\vrule}
    c c c c c c c
    !{\color{gray!40}\vrule}
    c c c c c c c
}
\toprule
&
\multicolumn{7}{c!{\color{gray!40}\vrule}}{\cellcolor{groupA}\textbf{Avg}} &
\multicolumn{7}{c!{\color{gray!40}\vrule}}{\cellcolor{groupB}\textbf{Vocals}} &
\multicolumn{7}{c!{\color{gray!40}\vrule}}{\cellcolor{groupA}\textbf{Bass}} &
\multicolumn{7}{c!{\color{gray!40}\vrule}}{\cellcolor{groupB}\textbf{Drums}} &
\multicolumn{7}{c}{\cellcolor{groupA}\textbf{Other}} \\
\cmidrule(lr){2-8} \cmidrule(lr){9-15} \cmidrule(lr){16-22} \cmidrule(lr){23-29} \cmidrule(lr){30-36}
\textbf{Method} &
\rotatebox{75}{\textbf{Mel}$\downarrow$} &
\rotatebox{75}{\textbf{SDR}$\uparrow$} &
\rotatebox{75}{\textbf{FD\textsubscript{VGG}}$\downarrow$} &
\rotatebox{75}{\textbf{FD\textsubscript{PaSST}}$\downarrow$} &
\rotatebox{75}{\textbf{FD\textsubscript{CLAP}}$\downarrow$} &
\rotatebox{75}{\textbf{FD\textsubscript{MERT}}$\downarrow$} &
\rotatebox{75}{\textbf{LPAPS\textsubscript{CLAP}}$\downarrow$} &
\rotatebox{75}{\textbf{Mel}$\downarrow$} &
\rotatebox{75}{\textbf{SDR}$\uparrow$} &
\rotatebox{75}{\textbf{FD\textsubscript{VGG}}$\downarrow$} &
\rotatebox{75}{\textbf{FD\textsubscript{PaSST}}$\downarrow$} &
\rotatebox{75}{\textbf{FD\textsubscript{CLAP}}$\downarrow$} &
\rotatebox{75}{\textbf{FD\textsubscript{MERT}}$\downarrow$} &
\rotatebox{75}{\textbf{LPAPS\textsubscript{CLAP}}$\downarrow$} &
\rotatebox{75}{\textbf{Mel}$\downarrow$} &
\rotatebox{75}{\textbf{SDR}$\uparrow$} &
\rotatebox{75}{\textbf{FD\textsubscript{VGG}}$\downarrow$} &
\rotatebox{75}{\textbf{FD\textsubscript{PaSST}}$\downarrow$} &
\rotatebox{75}{\textbf{FD\textsubscript{CLAP}}$\downarrow$} &
\rotatebox{75}{\textbf{FD\textsubscript{MERT}}$\downarrow$} &
\rotatebox{75}{\textbf{LPAPS\textsubscript{CLAP}}$\downarrow$} &
\rotatebox{75}{\textbf{Mel}$\downarrow$} &
\rotatebox{75}{\textbf{SDR}$\uparrow$} &
\rotatebox{75}{\textbf{FD\textsubscript{VGG}}$\downarrow$} &
\rotatebox{75}{\textbf{FD\textsubscript{PaSST}}$\downarrow$} &
\rotatebox{75}{\textbf{FD\textsubscript{CLAP}}$\downarrow$} &
\rotatebox{75}{\textbf{FD\textsubscript{MERT}}$\downarrow$} &
\rotatebox{75}{\textbf{LPAPS\textsubscript{CLAP}}$\downarrow$} &
\rotatebox{75}{\textbf{Mel}$\downarrow$} &
\rotatebox{75}{\textbf{SDR}$\uparrow$} &
\rotatebox{75}{\textbf{FD\textsubscript{VGG}}$\downarrow$} &
\rotatebox{75}{\textbf{FD\textsubscript{PaSST}}$\downarrow$} &
\rotatebox{75}{\textbf{FD\textsubscript{CLAP}}$\downarrow$} &
\rotatebox{75}{\textbf{FD\textsubscript{MERT}}$\downarrow$} &
\rotatebox{75}{\textbf{LPAPS\textsubscript{CLAP}}$\downarrow$} \\
\midrule

\rowcolor{headercolor}
\multicolumn{36}{l}{\textit{Discriminative}} \\

\makecell[l]{\textit{SCNet} $\ddagger$} &
1.66 & \textbf{9.69} & 1.62 & 286 & \underline{0.25} & 6.67 & \textbf{2.24} &
1.73 & \textbf{10.9}  & 1.77 & 338 & 0.28 & 10.9  & \underline{2.23} &
2.14 & \textbf{9.42} & 2.59 & 394 & \underline{0.38} & 9.25 & 2.88 &
1.54 & \textbf{11.2} & 1.45 & 170 & 0.19 & \underline{4.33} & \textbf{1.95} &
\textbf{1.22} & \textbf{7.28} & \textbf{0.67} & \textbf{240} &
\textbf{0.15} & \textbf{2.14} & \textbf{1.87} \\

\makecell[l]{HTDemucs $\ddagger$} &
\underline{1.61} & \underline{8.14} & 2.05 & 279 & 0.26 & 6.08 & \underline{2.28} &
1.67 & \underline{8.56} & 1.77 & 288 & \underline{0.24} & 9.42 & 2.24 &
\underline{1.99} & \underline{8.51} & 2.93 & \underline{365} & 0.40 & 8.38 & 2.83 &
\textbf{1.45} & \underline{9.26} & 1.89 & 173 & 0.20 & \textbf{4.25} & \underline{1.98} &
1.36 & \underline{6.22} & 1.51 & 290 & \underline{0.22} &
\underline{2.25} & \underline{2.06} \\

\makecell[l]{\textit{Open-Unmix} $\ddagger$} &
2.36 & 5.19 & 2.45 & 436 & 0.45 & 6.93 & 2.97 &
2.37 & 6.09 & 2.20 & 486 & 0.38 & 9.95 & 2.98 &
2.65 & 4.61 & 3.24 & 591 & 0.57 & 9.50 & 3.09 &
2.41 & 6.39 & 2.84 & 332 & 0.47 & 5.23 & 3.03 &
2.02 & 3.69 & 1.54 & 336 & 0.37 & 3.03 & 2.79 \\

\rowcolor{flomssBlock}
\makecell[l]{\textit{\textbf{FLOMSS-S-Disc.}} $\ddagger$} &
2.30 & 4.95 & 2.55 & 308 & 0.33 & 6.23 & 2.58 &
2.00 & 6.88 & 2.02 & 220 & 0.26 & 6.41 & 2.34 &
2.60 & 4.03 & 3.36 & 457 & 0.43 & 9.36 & 2.79 &
2.51 & 5.42 & 3.30 & 193 & 0.28 & 5.60 & 2.54 &
2.07 & 3.46 & 1.54 & 364 & 0.35 & 3.57 & 2.66 \\

\midrule
\rowcolor{headercolor}
\multicolumn{36}{l}{\textit{Generative}} \\

\makecell[l]{MGE-LDM $\dagger$} &
-- & -- & -- & -- & -- & -- & -- &
3.86 & -1.26 & 3.02 & 436 & 0.54 & 11.3  & 4.79 &
3.89 & 0.19 & 2.38 & 371 & 0.43 & 9.38 & 3.19 &
2.03 & 0.11 & 0.67 & 243 & 0.44 & 6.81 & 3.16 &
-- & -- & -- & -- & -- & -- & -- \\

\makecell[l]{InstGlow $\S$} &
3.08 & 2.75 & 3.96 & 535 & 0.69 & 10.6  & 3.27 &
3.87 & 3.07 & 4.04 & 624 & 0.68 & 12.3  & 3.76 &
3.12 & 2.27 & 4.34 & 695 & 0.80 & 17.2  & 2.97 &
2.68 & 3.49 & 3.28 & 353 & 0.71 & 7.59 & 3.20 &
2.67 & 2.15 & 4.17 & 467 & 0.56 & 5.42 & 3.15 \\

\makecell[l]{SAM Audio $\ast$} &
-- & -- & -- & -- & -- & -- & -- &
2.47 & 2.75 & 2.64 & 425 & 0.28 & 10.0  & 2.86 &
2.04 & 4.92 & 3.40 & 396 & 0.44 & \underline{8.24} & \underline{2.70} &
1.61 & 5.68 & 2.89 & 215 & \textbf{0.14} & 4.79 & 2.14 &
-- & -- & -- & -- & -- & -- & -- \\

\rowcolor{flomssBlock}
\textbf{\textit{FLOMSS-S}} $\ddagger$ &
1.63 & 3.77 & \underline{0.87} & \underline{261} & 0.31 &
\underline{5.43} & 2.45 &
\underline{1.58} & 5.08 & \underline{0.42} & \underline{162} & 0.26 &
\underline{5.72} & 2.34 &
2.01 & 3.30 & \underline{1.27} & 395 & 0.43 & 8.35 & 2.86 &
1.51 & 4.41 & \underline{0.61} & \underline{155} & 0.21 & 4.85 & 2.14 &
1.41 & 2.28 & 1.18 & 332 & 0.36 & 2.81 & 2.46 \\

\rowcolor{flomssBlock}
\textbf{\textit{FLOMSS-M}} $\ddagger$ &
\textbf{1.55} & 4.34 & \textbf{0.65} & \textbf{205} & \textbf{0.24} &
\textbf{5.15} & 2.30 &
\textbf{1.42} & 6.21 & \textbf{0.32} & \textbf{131} & \textbf{0.21} &
\textbf{5.63} & \textbf{2.21} &
\textbf{1.98} & 3.84 & \textbf{0.99} & \textbf{287} & \textbf{0.31} &
\textbf{7.70} & \textbf{2.64} &
\underline{1.46} & 4.75 & \textbf{0.56} & \textbf{138} & \underline{0.17} &
4.73 & 2.06 &
\underline{1.34} & 2.54 & \underline{0.71} & \underline{265} &
0.26 & 2.52 & 2.30 \\

\bottomrule
\end{tabular}
\end{adjustbox}
\begin{tablenotes}
\scriptsize
\item \ddag\ 2-channel at 44.1\,kHz (SCNet, HTDemucs, Open-Unmix, FLOMSS); \\ \dag\ 1-channel at 16\,kHz (MGE-LDM); \S\ 1-channel at 22.05\,kHz (InstGlow); $\ast$\ 1-channel at 48\,kHz (SAM Audio). 
\end{tablenotes}
\vspace{-3pt}

\vspace{-2mm}
\end{table*}

\begin{table}[t]
\centering
\vspace{-2.5mm}
\footnotesize
\renewcommand{\arraystretch}{1.02}
\setlength{\tabcolsep}{2.6pt}

\caption{Mean subjective scores are reported across individual sources.
Overall ranks exclude reference conditions and systems without results
for all four sources.}
\label{tab_subjective}

\begin{adjustbox}{max width=\columnwidth}
\begin{tabular}{
    l c
    !{\color{gray!40}\vrule}
    c c
    !{\color{gray!40}\vrule}
    c c c c
}
\toprule
\multicolumn{1}{c}{\textbf{Method}} &
\multicolumn{1}{c!{\color{gray!40}\vrule}}{\textbf{\# Params (M)}} &
\multicolumn{2}{c!{\color{gray!40}\vrule}}{\textbf{Overall}} &
\multicolumn{4}{c}{\textbf{Per-source}} \\
\cmidrule(lr){3-4}
\cmidrule(lr){5-8}
& &
\textbf{Avg.$\uparrow$} &
\textbf{Rank$\downarrow$} &
\textbf{Vocals$\uparrow$} &
\textbf{Bass$\uparrow$} &
\textbf{Drums$\uparrow$} &
\textbf{Other$\uparrow$} \\
\midrule

GT
& --
& 94.57 & -- 
& 96.64 & 92.90 & 94.33 & 94.40 \\

16\,kHz GT
& --
& 78.24 & -- 
& 73.62 & 89.29 & 67.40 & 82.65 \\

\midrule
\rowcolor{headercolor}
\multicolumn{8}{l}{\textit{Discriminative}} \\

\textit{SCNet} $\ddagger$
& 41
& \textbf{62.59} & \textbf{1} 
& \textbf{73.10}
& \underline{48.88}
& \underline{62.39}
& \textbf{65.98} \\

HTDemucs $\ddagger$
& 42
& 58.71 & 3 
& 65.00
& 43.38
& 60.82
& \underline{65.63} \\

\textit{Open-Unmix} $\ddagger$
& 36
& 35.40 & 6 
& 44.05
& 24.14
& 31.35
& 42.07 \\

\rowcolor{flomssBlock}
\textit{\textbf{FLOMSS-S-Disc.}} $\ddagger$
& 15 $\times$ 4 $=$ 60
& 40.28 & 5 
& 48.41
& 32.62
& 37.46
& 42.61 \\

\midrule
\rowcolor{headercolor}
\multicolumn{8}{l}{\textit{Generative}} \\

MGE-LDM $\dagger$
& 1,179
& -- & -- 
& 30.90
& 37.76
& 42.11
& -- \\

InstGlow $\S$
& 15 $\times$ 4 $=$ 60
& 26.6 & 7 
& 26.95
& 20.22
& 30.60
& 28.75 \\

SAM Audio $\ast$
& 709
& -- & -- 
& 45.60
& 41.14
& \textbf{66.54}
& -- \\

\rowcolor{flomssBlock}
\textbf{\textit{FLOMSS-S}} $\ddagger$
& 16 $\times$ 4 $=$ 64
& 53.99 & 4 
& 55.31
& 47.36
& 59.02
& 54.28 \\

\rowcolor{flomssBlock}
\textbf{\textit{FLOMSS-M}} $\ddagger$
& 67 $\times$ 4 $=$ 268
& \underline{59.12}
& \underline{\textbf{2}}
& \underline{67.16}
& \textbf{51.71}
& 61.21
& 56.40 \\

\bottomrule
\end{tabular}
\end{adjustbox}
\vspace{-2mm}

\vspace{-3mm}
\end{table}

\vspace{-1mm}
\section{Experiments}
\label{sec:experiments}
\vspace{-1mm}
\subsection{Experimental Setup}
\label{sec:exp_setup}

Unless otherwise stated, all experiments are based on a 16M-parameter NCSN++ U-Net~\cite{song2021scorebasedgenerativemodelingstochastic} backbone configured following SGMSE~\cite{Richter_2023}. 
To study model scaling, we increase the channel width to obtain a 67M-parameter model.
We condition the backbone by concatenating $\mathbf{x_t}$ and $\widetilde M$ at the input and separately providing $t$; the discriminative baseline receives $\widetilde M$ alone.

We compute a complex STFT with a Hann window size of 1534 and a hop length of 384
and compress the feature using $\alpha=0.5$ and $\beta=0.15$~\cite{Richter_2023}. For OT-CFM, we use $\sigma_{\max}=0.5$ and $\sigma_{\min}=0$. For SB-CFM, we set the Brownian-bridge noise scale $\sigma_\rho=1$, yielding a maximum standard deviation of \(0.5\) at \(t=0.5\). At inference, we use 10-step Midpoint ODE solver for OT-CFM and 10-step exponential integrator~\cite{he2024consistencydiffusionbridgemodels} for SB-CFM.

We use the official MUSDB18-HQ~\cite{Rafii2019MUSDB18HQA} split at 44.1 kHz stereo format. During training, we randomly crop audio segments with sample size of 97,920 and apply online remix augmentation: 
for each segment, we sample 2--4 stems from the same time region of the same song and remix them to form a new mixture. At inference, we process segments of the same sample size as training with 10\% overlap and combine them using triangular crossfades.
We train every model for 93,600 steps using Adam~\cite{kingma2017adammethodstochasticoptimization}, learning rate of $10^{-4}$, and effective batch size of 32 on 4 A100 GPUs.

\vspace{-2mm}
\subsection{Evaluation Metrics}
\label{ssec:evaluation-metrics}
\vspace{-2mm}
We use a collection of different metrics drawn from recent evaluations of audio generative models~\cite{comunita2024specmaskgitmaskedgenerativemodeling, zhong2025specmaskfoleysteeringpretrainedspectral, saito2025soundctmunifyingscorebasedconsistency, evans2024stableaudioopen}:\\
\noindent\textbf{Signal-based metrics.} We use spectral (STFT and Mel) and waveform L1 distance following DAC~\cite{kumar2023highfidelityaudiocompressionimproved}. 
We further report SDR using the MDX23 implementation~\cite{Fabbro_2024}.\\
\noindent \textbf{Perceptual similarity.} All embedding-based metrics are computed using non-overlapping
10-s audio segments extracted from the estimated sources and their corresponding ground truth. We compute LPAPS~\cite{iashin2021tamingvisuallyguidedsound} with CLAP embeddings~\cite{wu2024largescalecontrastivelanguageaudiopretraining} between corresponding 10-s segments of the estimated sources
and ground truth.\\
\noindent\textbf{Distributional metrics.} Different from LPAPS, which computes the pairwise distance, distribution matching metrics compare the sets of embeddings extracted from all 10-s ground-truth and separated segments. We report widely adopted Fr\'echet Distance (FD)~\cite{kilgour2019frechetaudiodistancemetric}, and Maximum Mean Discrepancy (MMD)~\cite{Gretton2012AKT} using VGGish-16k~\cite{hershey2017cnnarchitectureslargescaleaudio}, PaSST-32k~\cite{Koutini_2022}, OpenL3-48k~\cite{8682475}, CLAP-48k~\cite{wu2024largescalecontrastivelanguageaudiopretraining}, and MERT-24kHz~\cite{Li2023MERTAM}.

\noindent\textbf{Subjective quality.} We conduct a listening test with WebMUSHRA \cite{schoeffler2018webmushra}. 23 participants with different levels of experience in music production participated in the test. Each participant rated the overall quality of ten 8-second excerpts on a 0--100 scale. Conditions are anonymized and randomly ordered. We include the full-band ground-truth and its 16-kHz version as a bandwidth-limited hidden reference for generative MSS baselines that do not support full-band audio. For each method, we report the mean score across all participants and evaluated excerpts.
\vspace{-3.5mm}
\section{Results}
\vspace{-2mm}
\subsection{Ablation Study with the Single-source Modeling}\label{5.1}
Table~\ref{tab_ablation_all_stems} compares the CFM formulations with the discriminative baseline and unprocessed mixture.
The discriminative baseline achieves the best waveform distance and SDR, whereas both OT-CFM variants reduce the Mel and STFT distances and improve most embedding-based metrics. Among the CFM variants, SB-CFM-$\mathbf{x}$ performs best on waveform distance and SDR but consistently underperforms the OT-CFM variants on embedding-based metrics.
Comparing the OT-CFM parameterizations, $\mathbf{v}$-prediction achieves the best scores in embedding-based metrics, but yields worse SDR; whereas $\mathbf{x}$-prediction performs better on all signal-based metrics. 

We use OT-CFM-$\mathbf{x}$ for subsequent experiments because of its balanced performance across our collection of metrics.

\vspace{-2mm}
\subsection{Stepwise Behaviors}
\vspace{-1mm}




Fig.~\ref{fig:midpoint_sampler} compares the intermediate predictions of single-source model for Vocals using OT-CFM-$\mathbf{x}$, with the discriminative baseline shown by dashed lines. 
Early target estimates approach the discriminative baseline in Mel distance and SDR, suggesting that the model first recovers a reconstruction-oriented estimate that is closer to its discriminative version; 
later estimates generally improve spectral and embedding-based metrics by sacrificing the SDR score, revealing a trade-off between spectral \& embedding-based similarity and waveform reconstruction fidelity. 
For the subsequent analyses, we retain all embedding backbones except OpenL3 and use FD as the representative distributional metric. Complete results for all evaluated metrics will be provided in our open-source repository.
\vspace{-2mm}
\subsection{Multi-Source Modeling}
\vspace{-1mm}
Table~\ref{tab_modeling} compares single-source OT-CFM-$\mathbf{x}$ with its joint four-source variants. 
The unconstrained multi-source model ( denoted as ``w/o MC'') leaves the backbone size nearly unchanged and produces all four stems within a single model, rather than using four different single-source models. This efficiency in turn harms its separation quality, particularly on embedding-based metrics. 

Adding the soft MC constraint described in Sec.~\ref{ssec:multi-source modeling} recovers most of the SDR loss, while preserving Mel and LPAPS. 
Nevertheless, a hard MC constraint, either in the waveform domain or in the uncompressed STFT domain, exhibits an overall degradation, except with MERT embeddings.
Unlike \cite{scheibler2025sourceseparationflowmatching}, we omit auxiliary signal-level losses to keep the training objectives consistent in our ablation study. This finding is limited to the waveform- and STFT-domain paths evaluated in this study and suggests that hard MC may require an MSS-specific flow path and training strategy. We leave this as our future work.

\vspace{-3mm}
\subsection{Benchmarking with Prior Methods}
\vspace{-1mm}
In Table~\ref{tab_benchmark}, we benchmark our FLOMSS (single-source OT-CFM-$\mathbf{x}$) with existing generative and discriminative MSS methods. We note that these comparisons should be interpreted with care: SAM Audio and MGE-LDM were trained on different datasets, HTDemucs~\cite{rouard2022hybridtransformersmusicsource} was trained on MUSDB18-HQ augmented with 800 extra songs, InstGlow was trained on MUSDB18 augmented with guitar and piano stems from Slakh2100~\cite{manilow2019cutting}, while SCNet~\cite{Tong2024SCNetSC}, Open-Unmix~\cite{stoter19} and our models use MUSDB18-HQ training split.

In Table~\ref{tab_benchmark}, FLOMSS-S (16M) already outperforms prior generative baselines on most spectral and perceptual metrics under format-matched evaluation. 
Compared with FLOMSS-S-Disc. (same-backbone discriminative baseline), it improves Mel and all embedding-based metrics, while yielding a slightly lower SDR score, consistent with the observed distributional–reconstruction trade-off in Table~\ref{tab_ablation_all_stems}. 
Scaling to FLOMSS-M (67M) improves every metric and surpasses Open-Unmix on all metrics except SDR. Compared with SCNet and HTDemucs, FLOMSS-M achieves better Mel and FD scores on averages across all embeddings, while retaining a weaker SDR and slightly worse LPAPS.
We leave the further enhancement in SDR as the future work.

Last but not least, Table~\ref{tab_subjective} presents the subjective quality of these MSS methods via a listening test described in Sec.~\ref{ssec:evaluation-metrics}. Despite its lower SDR, FLOMSS-M ranked no.2 in the overall score, and ranked no.1 in Bass, no.2 in Vocals, and no.3 in Drums, indicating competitive perceptual quality.





\vspace{-2mm}
\section{Conclusion}
\vspace{-1mm}
We present FLOMSS, an open-source flow-matching framework for full-band stereo MSS, together with a systematic study of prediction targets, flow paths, single- and multi-source modeling, and mixture consistency. By extending evaluation beyond SDR to spectral and embedding-based metrics, we identify a distributional–reconstruction trade-off in generative MSS. On MUSDB18-HQ, FLOMSS surpasses prior generative methods and a same-backbone discriminative baseline on spectral and embedding-based metrics. It also improves Mel and FD over strong discriminative methods, although SDR remains lower. These results establish flow matching as a promising direction for generative MSS.

\newpage
\bibliographystyle{IEEEbib}
{
\footnotesize
\bibliography{refs}
}
\end{document}